\parindent=0pt
\parskip=0.3cm
\magnification \magstephalf

\null
\vskip 1.5cm
\centerline{{\bf APOCALYPSE SOON}
\footnote{\dag}{Mon. Not. R. Astron. Soc. 296 (1998) 619}}
\vskip 1.5cm

\centerline {by}
\vskip 0.3cm

\centerline {J.C. Jackson\footnote{*}{e-mail: john.jackson@unn.ac.uk}}
\centerline {Department of Mathematics and Statistics}
\centerline {University of Northumbria at Newcastle}
\centerline {Ellison Building}
\centerline {Newcastle upon Tyne NE1 8ST, UK}
\vskip 1.5cm

\centerline{\bf ABSTRACT}
\vskip 0.3cm
Based upon a simple vacuum Lagrangian, comprising cosmological and 
quadratic scalar field terms, a cosmological model is presented whose 
history is indistinguishable from that of an innocuous low-density 
cold dark matter (CDM) universe, but whose future is very much shorter.  
For sensible values of the deceleration parameter ($0<q_0<1$), its age 
is greater than 85\% of the Hubble time, thus resolving the current 
version of the age crisis, which appears to be that $t_0\sim1/H_0$ 
while $q_0$ is significantly positive.
\vskip 0.6cm

{\bf Key words:} cosmology -- observations -- theory -- dark matter. 
\vfil\eject

{\bf 1 INTRODUCTION}

That there is something wrong with our understanding of the dynamics of 
large astronomical systems, characterised by length scales ranging from
10 kpc (galaxies) to 100 Mpc (deviations from the Hubble flow) seems to
be beyond doubt, which problems have been `solved' by the introduction
of dark matter.  Beyond 100 Mpc, there are indications that the Hubble
expansion is slowing down, (Perlmutter et al. 1996, 1997a,b), 
to an extent which would also require significant amounts of dark matter, 
corresponding to a deceleration parameter $q_0\sim 1/2$.  

I have long had an interest in the idea that the solution to the 
mystery of extra-galactic dynamics might be found in the very 
fabric of space-time (i.e. the vacuum), rather than its contents 
(Jackson 1970, 1992; Jackson \& Dodgson 1996, 1997).  Traditionally 
the concept of vacuum energy finds its expression in the form of a 
cosmological term in Einstein's equations, which in effect assigns an 
energy-density $\lambda/8\pi G$ to the vacuum.  We have argued 
(Jackson \& Dodgson 1998) that the natural generalization of this concept 
introduces in addition a scalar field $\phi$, with corresponding Lagrangian
${\cal L}=\dot\phi^2/2-V(\phi)$, where the potential $V(\phi)$ 
has a minimum at $\phi=0$.  The energy-density attributed to the vacuum
is now $\dot\phi^2/2+V(\phi)$, and in the proposed model $V(\phi)$
has the simple form $\lambda/8\pi G+\omega_c^2\phi^2/2$, 
which is generic in the sense that it is not tied to a particular 
field-theoretic model, but is just a Maclaurin expansion in which only 
the first three terms are retained; $\omega_c$ is the Compton frequency 
of an associated ultra-light boson (cf. Frieman et al. 1995).  Thus
$$
\rho_{VAC}=\dot\phi^2/2+\lambda/8\pi G+\omega_c^2\phi^2/2. 
\eqno(1)
$$
This is of course the language of chaotic inflationary cosmology
(Linde 1983), but the point here is that with the appropriate value
of $\omega_c$ (i.e. $\omega_c^{-1}{}^{>}_\sim$ current Hubble time, rather 
than a tiny fraction of a second), `inflation' is something which might 
have happened in the recent past, rather than a long time ago
(Frieman et al. 1995; Jackson \& Dodgson 1998).
In addition to $\rho_{VAC}$ a real Universe would require a trace
of more conventional matter, Cold Dark Matter plus baryonic, or possibly 
just the observed baryons; there would be no technical difficulties 
associated with such a component, but in what follows I shall neglect 
its marginal dynamical effects, the inclusion of which would serve only to 
obscure the main points of this communication.  The general view adopted 
here is that ordinary matter is just flotsam floating on a deep and 
possibly rough scalar sea.

Our initial motivation in this context (Jackson \& Dodgson 1998) was 
occasioned by the recent upward trend in measures of $H_0$ 
(Pierce et al. 1994; Freedman et al. 1994; Tanvir et al. 1995), 
when the observational evidence seemed to indicate that $t_0>1/H_0$, 
and that $q_0>0$, which behaviour is allowed by the class of models 
described above; the scalar field mimics a positive cosmological constant 
in an inflationary slow-roll phase until the Hubble time exceeds 
$\omega_c^{-1}$, after which deceleration commences. 
The upward trend in $H_0$ values has been reversed somewhat by data from 
the Hipparcos astrometry satellite (Feast \& Catchpole 1997; see, however, 
Madore \& Freedman 1998), and the age crisis may not be as severe
as it seemed a few years ago; the current view is probably that 
$t_0\sim 1/H_0$, corresponding for example to a low-density CDM model,
but significantly longer than the value $2/3\times 1/H_0$ demanded by the 
canonical flat CDM model.  However, quite apart from these considerations, 
I believe that the class of models described by equation (1) deserves 
further study, and I present here one which seems to me to have some 
particularly interesting properties.
\vskip 0.6cm

{\bf 2 STAGFLATION}

A peculiar feature of the class is that it offers two mechanisms for
generating a cosmological constant, the necessarily positive slow-roll 
($\dot\phi<<\omega_c\phi$) inflationary term $4\pi G\omega_c^2\phi^2$, 
and the conventional $\lambda=8\pi GV(0)$ deriving from a non-zero 
minimum value of $V$.  There seems to be little point in retaining the 
conventional $\lambda$ term as an extra positive contribution, but 
interesting variations in behaviour arise if $V(0)$, and hence $\lambda$, 
is negative; a possibility which has particularly captured my imagination
is a slow-roll expanding phase in which $\phi=\phi_0$ initially and the 
last two terms in equation (1) cancel: 
$$
\lambda=-4\pi G\omega_c^2\phi_0^2, 
\eqno(2)
$$
which might be called stagflation. 

The scalar field has effective density $\rho_\phi=
(\dot\phi^2+\omega_c^2\phi^2)/2$ and pressure 
$p_\phi=(\dot\phi^2-\omega_c^2\phi^2)/2$, and for a model in which these 
are dominant, the Friedmann equations for the scale factor $R(t)$ are
$$
\ddot R=-{4\pi \over 3}G(\rho_\phi+3p_\phi)R+{1 \over 3}\lambda R=
-{4\pi \over 3}G(2\dot\phi^2-\omega_c^2\phi^2)R+{1 \over 3}\lambda R,
\eqno(3)
$$
$$
\dot R ^2={8\pi \over 3}G\rho_\phi R^2+{1 \over 3}\lambda R^2-kc^2.
\eqno(4)
$$
The scalar field is governed by the equation
$$
\ddot\phi+3H\dot\phi+\omega_c^2\phi=0,
\eqno(5)
$$
where $H=\dot R/R$ is Hubble's `constant' (see for example Peebles 1993).

During the slow-roll phase we have $\dot\phi<<\omega_c\phi$, and thus with
$\lambda=-4\pi G\omega_c^2\phi_0^2$ equation (3) gives $\ddot R\sim 0$, 
which brings us to the first interesting feature of the model: 

{\parindent 0.5cm
\item{i)} its history is indistinguishable from that 
of an innocuous low-density CDM universe, but its future might be 
very different.
}

It is possible to trace analytically the beginnings of said future, 
which is increasingly dominated by an imbalance as $\phi$ drifts away 
from the value indicated by equation (2).  During the slow-roll phase,
we have $H=1/t$, and equation (5) becomes
$$
\ddot\phi+{3 \over t}\dot\phi+\omega_c^2\phi=0.
\eqno(6)
$$
An exponential inflationary phase is characterised by a constant value of $H$, 
when the $t$ in the second term of equation (6) has a fixed value $t_H$,
and $\ddot\phi<<3\dot\phi/t_H$, so that the first term in equation (6)
can be neglected.  This is not the case here, but a non-singular series
solution for $\phi$ is easily developed, to give
$$
\phi=\phi_0(1-\omega_c^2t^2/8+\omega_c^4t^4/192-......),
\eqno(7)
$$
To order $\omega_c^2t^2$, the scalar density and pressure are now 
$$
\rho_\phi
={\omega_c^2\phi_0^2 \over 2}\left(1-{3 \over 16}\omega_c^2t^2\right),
~~~~~~~~~~~~ 
p_\phi
=-{\omega_c^2\phi_0^2 \over 2}\left(1-{5 \over 16}\omega_c^2t^2\right), 
\eqno(8)
$$
and equation (3) becomes
$$
\ddot R
=-{1 \over 8}(4\pi G\omega_c^2\phi_0^2)\omega_c^2t^2R
=-{1 \over 8}|\lambda|\omega_c^2t^2R,
\eqno(9)
$$
or, in terms of the deceleration parameter $q=-\ddot RR/\dot R^2$, 
$$
q
={1 \over 8}|\lambda|\omega_c^2t^4\Rightarrow t(q)
={\left(8q \over |\lambda|\omega_c^2\right)}^{1/4}.
\eqno(10)
$$
Thus it turns out that, for fixed $q_0$ and $t_0$, $|\lambda|$ and hence 
$\phi_0$ can be arbitrarily large, as long as $\omega_c$ is appropriately 
small: 
$$
\omega_c=\left({8q_0 \over |\lambda|t_0^4}\right)^{1/2}.  
\eqno(11)
$$
However, to allow an exit from the slow-roll phase, and the possibility
that $q_0\not=0$, $\omega_c$ cannot be exactly zero.  The second
attractive feature of the model is thus:

{\parindent=0.5cm
\item{ii)} it serves as a paradigm for what might be happening in 
the real vacuum, in which the huge zero-point energy suggested by 
quantum field theory is renormalized by a counter term.  
}

To this order of approximation, equation (9) integrates to give the
scale factor explicitly as
$$
R(t)=t\left(1-{|\lambda|\omega_c^2t^4 \over 160}\right).
\eqno(12)
$$
It is clear that evolution is determined by the composite parameter
$|\lambda|\omega_c^2$, and that only this combination can be fixed 
by cosmological observations, for example via equation (10) from 
knowledge of $q_0$ and $t_0$.  Equation(4) gives $\dot R^2=-kc^2$
initially, so that $k=-1$ and the model is an open one.

To follow the evolution in full, equations (3) to (5) must be
integrated numerically, which is best achieved by using the above
analytical solution to give a convenient starting point which avoids 
the initial singularity.  Figure 1 shows the resulting trajectory, 
normalized in terms of both $R$ and $t$ to the point where $q_0=1/2$,
which curve is universal in the sense that it does not depend upon the
particular value of $|\lambda|\omega_c^2$, in the the limit of large
values of this parameter.  Figure 1 illustrates a third feature of this 
universe:

{\parindent=0.6cm
\item{iii)} for all conceivable values of the deceleration parameter,
its age is greater than 70\% of the Hubble time,
}

which aspect, from the point of practical cosmology, is probably its most
important.  This behaviour is quantified in Figure 2, which shows $Ht$ as 
a function of $q$, together with the conventional CDM curve for comparison;
thus for example $q_0=1/2$ corresponds to $H_0t_0=0.91$ here, much greater 
than the CDM figure of 0.67.  There is however a terrible price to pay for an 
extended history; violent re-collapse is inevitable, and the future is very 
much shorter than might be expected; the final stages of the collapse 
are in fact dominated by the $\dot\phi^2$ terms in the expressions for 
$\rho_\phi$ and $p_\phi$, when $p_\phi\sim +\rho_\phi$, and the scalar 
terms in equation (3) reinforce the effects of the negative cosmological 
constant.  The title of this note is explained by such behaviour.

In writing this letter I am well aware of the possibility that a reworking 
of stellar evolution theory, coupled with the recent downturn in measures of 
$H_0$, may re-establish a more conventional view of cosmic history.  
A putative cosmological model must expect to run the gauntlet of an
increasing number of neo-classical observational tests ($z{}^{<}_\sim 4$), 
and of modern tests based upon structure formation and the cosmic microwave
background ($z{}^{<}_\sim 1000$), but in the context of this model I would  
not be prepared to further action until such time as the age crisis 
has a firmer observational basis.  The crisis may take the extreme form 
$t_0>1/H_0$, $q_0>0$ discussed by Jackson \& Dodgson (1998), or the more 
modest form discussed here, namely $t_0{}^{<}_\sim 1/H_0$ but too long to be 
compatible with significant deceleration.  Radical measures would
seem to be unavoidable, if either possibility is confirmed.
\vskip 0.6cm

{\bf REFERENCES}
\vskip 0.3cm

{\obeylines\parskip=0pt
Feast M.W., Catchpole R.M., 1997, MNRAS, 286, L1
Freedman, W.L. et al., 1994, Nat, 371, 757 
Frieman J.A., Hill C.T., Stebbins A., Waga I., 1995, Phys. Rev. Lett., 75, 2077

Jackson J.C., 1970, MNRAS, 148, 249
Jackson J.C., 1992, QJRAS, 33, 17-26
Jackson J.C., Dodgson M., 1996, MNRAS, 278, 603
Jackson J.C., Dodgson M., 1997, MNRAS, 285, 806
Jackson J.C., Dodgson M., 1998, MNRAS, in the press

Linde A.D., 1983, Phys. Lett., B129, 177

Madore B.F., Freedman W.L., 1998, ApJ, 492, 110

Peebles P.J.E., 1993, Principles of Physical Cosmology.  Princeton University Press, pp. 394-396
Perlmutter S. et al., 1996, Nucl. Phys., S51B, 20
Perlmutter S. et al., 1997a, in Canal R., Ruiz-LaPuente P., Isern J., eds., 
\qquad Thermonuclear Supernovae, Kluwer, Dordrecht, pp. 749-763
Perlmutter S. et al., 1997b, ApJ, 483, 565
Pierce M.J., Welch D.L., McClure R.D., van den Bergh S., Racine R., Stetson P.B.,
\qquad 1994, Nat, 371, 385

Tanvir N.R., Shanks, T., Ferguson H.C., Robinson D.R.T., 1995, Nat, 377, 27
}
\vskip 0.6cm

{\bf FIGURE CAPTIONS}
\vskip 0.3cm

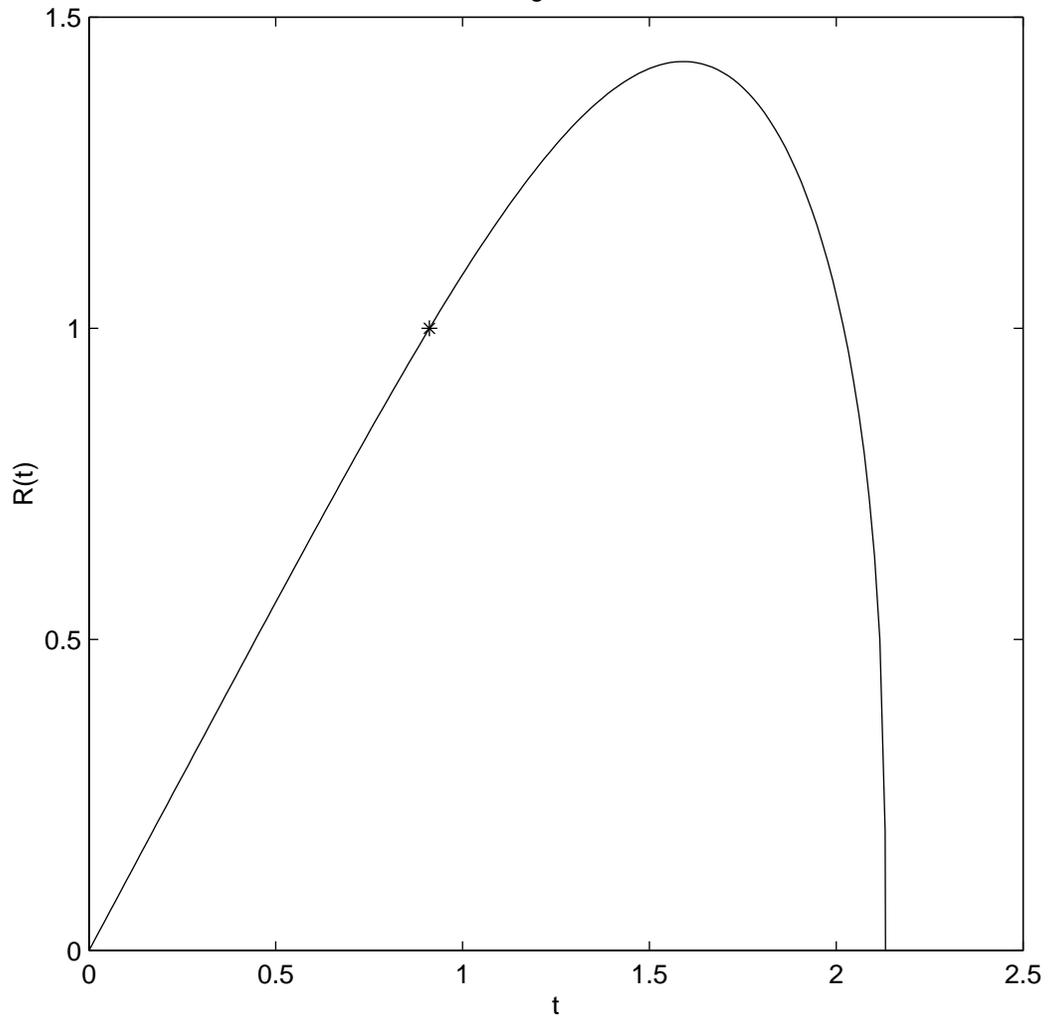
Figure 1.  Scale factor $R(t)$ for the stagflationary model; the asterisk 
marks the point at which $q=0.5$. 
\vskip 0.3cm

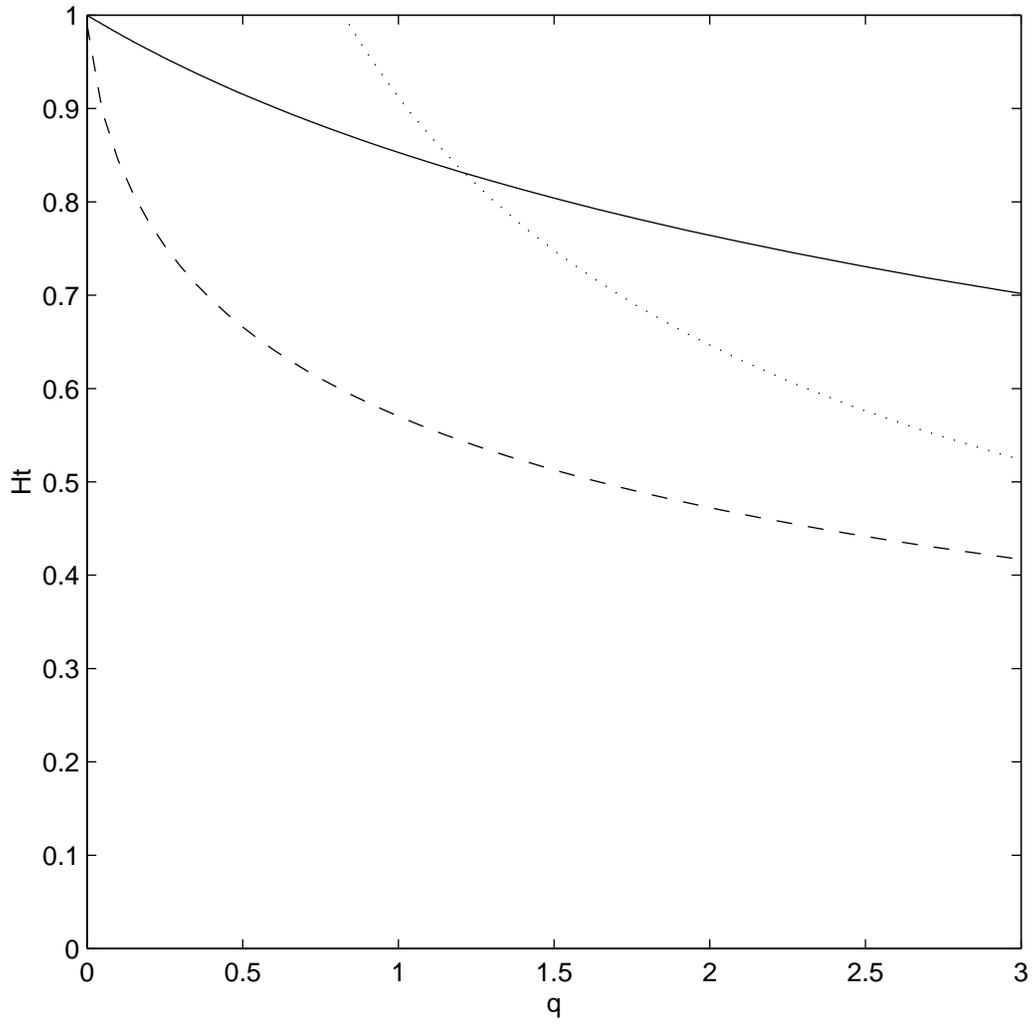
Figure 2.  The continuous line shows the age of the stagflationary 
model in Hubble units, as a function of the deceleration parameter $q$; 
the dashed line is the corresponding CDM curve.  The dotted line shows 
the time remaining in the stagflationary case; the corresponding CDM
times are too long to appear here, being $Ht\geq 1.27$ for $q\leq3$, and
of course infinite for $q\leq 1/2$.

\bye